\documentclass[12pt]{article}
\usepackage[dvips]{color}
\usepackage{epsfig}
\usepackage{amsmath}
\usepackage{graphicx}

\begin{document}
\title{\bf Dark Energy and Tachyon Field in Bianchi Type-V
Space-time}
\author{\small{J. Sadeghi$^{a}$\thanks{Email: pouriya@ipm.ir}, and H. Farahani$^{a}$\thanks{Email: h.farahani@umz.ac.ir}}\\
$^{a}${\small {\em Department of Physics, Mazandaran University,
Babolsar, Iran}}}  \maketitle
\begin{abstract}
In this paper, we consider Bianchi type-V
space-time and study a cosmological model of dark energy based on Tachyon scalar field. We assumed three different kinds of matter without possibility of interaction with scalar dark energy. Assuming power law Hubble parameter in terms of scale factor we obtain evolution of scalar field, scalar potential and equation of state parameter.
\end{abstract}

\section{Introduction}
Recent cosmological observations [1-5] indicated that the universe expand with acceleration. Such accelerated expansion may described by dark energy models. The simplest candidate for the dark energy is the cosmological constant or vacuum energy which has good agreement with the observational data. However, many recently it is claimed that the cosmological constant may not be vacuum energy [6]. Moreover, the cosmological constant model has two famous problems as the fine-tuning and cosmic coincidence [7]. Therefore another models of the dark energy proposed include
quintessence [8, 9, 10], phantom [11, 12] and quintom [13, 14, 15]. Tachyon dark energy is also interesting model based on scalar field [16]. The tachyon scalar field was proposed as a source of the dark energy and inflation. The tachyon dark energy has EoS parameter between -1 and 0 [17]. The tachyon
is a an unstable field which has became important in string theory through its role in the
Dirac-Born-Infeld action which is used to describe the D-brane action [18].\\
There is also another interesting model to describe dark energy based on Chaplygin gas (CG) equation of state [19] which has several extensions such as generalized Chaplygin gas (GCG) [20, 21, 22], modified Chaplygin gas (MCG) [23, 24], modified cosmic Chaplygin gas (MCCG) [25, 26, 27], holographic Chaplygin gas [28, 29] and recently proposed extended Chaplygin gas (ECG) [30].\\
The anisotropy of the dark energy within the framework of Bianchi type space-times is found to be useful in generating
arbitrary ellipsoidality to the universe [31]. Bianchi type-V universe is generalization of the open Universe in FRW cosmology, therefore it is important to study the dark energy models in universe with non-zero curvature. Already Bianchi type-V model in different physical contexts have studied [32, 33, 34]. In the Ref. [35] isotropic dark energy model with variable EoS parameter in Bianchi type V space-time have studied and found that the present universe is dominated by the dark energy.\\
In this paper we consider Bianchi type-V universe and study Tachyon dark energy model. We also include matter contribution with several equation of states such as pressureless matter, barotropic matter, and modified Chaplygin gas. In the last model we assumed modified Chaplygin gas as a matter not dark energy. As we know, the Chaplygin gas plays the role of dark matter at early times, but at late times it tends to a cosmological constant.\\
This paper organized as follows. In next section we introduce our models, then in section 3 we write field equations and in section 4 try to solve them. Cosmological parameters analyzed in section 5 analytically and numerically. Stability of our models investigated in section 6. Finally in section 7 we give conclusion.

\section{The models}
In this paper, we consider a universe filled with dark energy and matter. For the dark energy we choose Tachyon field with the following energy density and pressure respectively [16],
\begin{equation}\label{s1}
p_{\phi}=-V(\phi)\sqrt{1-\dot{\phi}^{2}},
\end{equation}
and,
\begin{equation}\label{s2}
\rho_{\phi}=\frac{V(\phi)}{\sqrt{1-\dot{\phi}^{2}}}.
\end{equation}
Therefore, the equation of state of tachyon field obtained as follow,
\begin{equation}\label{s3}
\omega_{\phi}=\frac{p_{\phi}}{\rho_{\phi}}=\dot{\phi}^{2}-1.
\end{equation}
Also Tachyon potential is given by,
\begin{equation}\label{s4}
V(\phi)=\sqrt{-p_{\phi}\rho_{\phi}}.
\end{equation}
For the matter, we will consider three different models based on pressureless matter ($p=\omega=0$), barotropic matter with constant EoS parameter $\omega_{b}=p_{b}/\rho_{b}$, and modified Chaplygin gas with the following EoS parameter [36],
\begin{equation}\label{s5}
\omega_{mcg}=\frac{p_{mcg}}{\rho_{mcg}}=\gamma-\frac{\mathcal{B}}{\rho_{mcg}^{1+\alpha}},
\end{equation}
where $\gamma$ and $\mathcal{B}$ are positive constants.\\
We will consider three models of matter described above together Tachyon scalar field dark energy in Bianchi type-V
space-time as different models to describe universe. In the next section we review Bianchi type-V
space-time and field equations which govern our model.
\section{Field equations}
The spatially homogeneous and anisotropic Bianchi-V space-time is described by the following metric,
\begin{equation}\label{s6}
ds^2=-dt^2+A^{2}dx^{2}+e^{2\beta x}(B^{2}dy^{2}+C^{2}dz^{2}),
\end{equation}
where $A$, $B$ and $C$ are the metric functions of cosmic time $t$, and $\beta$ is a constant. The metric functions related to the scale factor via the following relation,
\begin{equation}\label{s7}
a=(ABC)^{\frac{1}{3}}.
\end{equation}
Therefore, Hubble expansion parameter is defined as the follow,
\begin{equation}\label{s8}
H=\frac{\dot{a}}{a}=\frac{1}{3}\left(\frac{\dot{A}}{A}+\frac{\dot{B}}{B}+\frac{\dot{C}}{C}\right),
\end{equation}
where an over dot denotes derivative with respect to the cosmic time $t$.\\
The Einstein's field equations, in the units $8\pi G=c=1$, read as,
\begin{equation}\label{s9}
R_{ij}-\frac{1}{2}g_{ij}R=-T_{ij},
\end{equation}
where $T_{ij} = T^{(m)}_{ij} +T^{(de)}_{ij}$ is the overall energy momentum tensor with $T^{(m)}_{ij}$ and $T^{(de)}_{ij}$ as the energy momentum
tensors of matter (pressureless matter, barotorpic matter or modified Chaplygin gas) and dark energy (Tachon field), respectively. We assume non-interacting models, hence, the energy conservation equations read as,
\begin{equation}\label{s10}
\dot{\rho}^{(m)}+3(1+\omega^{(m)})\rho^{(m)}H=0,
\end{equation}
and,
\begin{equation}\label{s11}
\dot{\rho}^{(de)}+3(1+\omega^{(de)})\rho^{(de)}H=0.
\end{equation}
In the next section we write general solution of the equation (11) for the dark energy density.
\section{General solutions}
According to the Ref. [34] we can use the following form of the Hubble expansion parameter,
\begin{equation}\label{s12}
H=Da^{-n},
\end{equation}
where $D$ is positive constant and $n\geq0$. Separate solutions obtained for $n=0$ and $n\neq0$ as follows [35].
\subsection{$n=0$}
In this case, from the equation (12), one can obtain the following scale factor,
\begin{equation}\label{s13}
a=c_{1}e^{Dt},
\end{equation}
where $c_{1}$ is an integration constant. Therefore, the dark energy density and pressure will be,
\begin{equation}\label{s14}
\rho^{(de)}=3D^{2}-\frac{l^{2}}{c_{1}^{6}}e^{-6Dt}-3\frac{\beta^{2}}{c_{1}^{2}}e^{-2Dt}-\rho^{(m)},
\end{equation}
and,
\begin{equation}\label{s15}
p^{(de)}=-3D^{2}-\frac{l^{2}}{c_{1}^{6}}e^{-6Dt}+\frac{\beta^{2}}{c_{1}^{2}}e^{-2Dt},
\end{equation}
where $l$ is an integration constant.
\subsection{$n\neq0$}
In that case one can obtain,
\begin{equation}\label{s16}
a=(nDt+c_{2})^{\frac{1}{n}},
\end{equation}
where $c_{2}$ is an integration constant. Therefore, the dark energy density and pressure will be,
\begin{equation}\label{s17}
\rho^{(de)}=\frac{3D^{2}}{(nDt+c_{2})^{2}}-\frac{l^{2}}{(nDt+c_{2})^{\frac{6}{n}}}
-3\frac{\beta^{2}}{(nDt+c_{2})^{\frac{2}{n}}}-\rho^{(m)},
\end{equation}
and,
\begin{equation}\label{s18}
p^{(de)}=\frac{(2n-3)D^{2}}{(nDt+c_{2})^{2}}-\frac{l^{2}}{(nDt+c_{2})^{\frac{6}{n}}}
+\frac{\beta^{2}}{(nDt+c_{2})^{\frac{2}{n}}}-\omega^{(m)}\rho^{(m)}.
\end{equation}
In order to avoid degeneracy we have a condition as $n\neq3$.\\
In the next section we will use above results to obtain cosmological parameter and discuss about Tachyon scalar field and potential.
\section{Cosmological parameters}
We will exam our model for three different kinds of matter. The simplest one is pressureless matter with $\omega^{(m)}=0$, which has straight generalization to barotropic matter with constant $\omega^{(m)}$. The last case, which is more complicated, is modified Chaplygin gas with $\omega^{(m)}$ given by the equation (5).
\subsection{Pressureless matter}

\subsubsection{$n=0$}
In that case one can obtain,
\begin{equation}\label{s19}
\rho^{(m)}=c_{0}e^{-3Dt},
\end{equation}
where $c_{0}$ is an arbitrary constant. Therefore, using the relations (3), (14), (15) and (19) we can obtain,
\begin{equation}\label{s20}
\dot{\phi}^{2}=\frac{2\frac{l^{2}}{c_{1}^{6}}e^{-6Dt}+2\frac{\beta^{2}}{c_{1}^{2}}e^{-2Dt}+c_{0}e^{-3Dt}}
{\frac{l^{2}}{c_{1}^{6}}e^{-6Dt}+3\frac{\beta^{2}}{c_{1}^{2}}e^{-2Dt}+c_{0}e^{-3Dt}-3D^{2}}.
\end{equation}
Then, using the equation (4) we find the Tachyon potential as follow,
\begin{eqnarray}\label{s21}
V^{2}=9D^{4}&-&12D^{2}\frac{\beta^{2}}{c_{1}^{2}}e^{-2Dt}-3D^{2}c_{0}e^{-3Dt}+3\frac{\beta^{4}}{c_{1}^{4}}e^{-4Dt}+c_{0}\frac{\beta^{2}}{c_{1}^{2}}e^{-5Dt}\nonumber\\
&-&2\frac{l^{2}\beta^{2}}{c_{1}^{8}}e^{-8Dt}-c_{0}\frac{l^{2}}{c_{1}^{6}}e^{-9Dt}-\frac{l^{4}}{c_{1}^{12}}e^{-12Dt}.
\end{eqnarray}
It is clear that the Tachyon potential yields to $3D^{2}$ at the late time.\\
In Fig. 1 we obtain behavior of the scalar field numerically. It is find that the scalar field is increasing function of time. We can see that variation of $D$ is not many important at the early time, after that, increasing $D$ increases value of $\phi$.\\

\begin{figure}[th]
\begin{center}
\includegraphics[scale=.35]{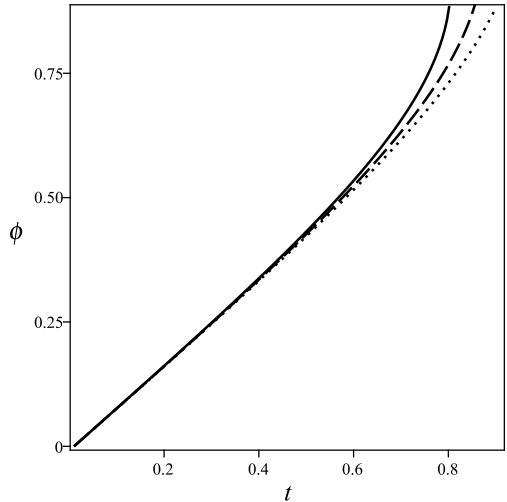}
\caption{Plot of scalar field in terms of cosmic time with $c_{0}=2$, $\frac{l^{2}}{c_{1}^{6}}=5$ and $\frac{\beta^{2}}{c_{1}^{2}}=100$. $D=2$ (solid line), $D=1.9$ (dashed line), $D=1.8$ (dotted line).}
\end{center}
\end{figure}

\subsubsection{$n\neq0$}
In that case one can obtain,
\begin{equation}\label{s22}
\rho^{(m)}=c_{0}(nDt+c_{2})^{-\frac{3}{n}},
\end{equation}
where $c_{0}$ is an arbitrary constant as before. Therefore, using the relations (3), (17), (18) and (22) we can obtain,
\begin{equation}\label{s23}
\dot{\phi}^{2}=\frac{\frac{2nD^{2}}{(nDt+c_{2})^{2}}-\frac{2l^{2}}{(nDt+c_{2})^{\frac{6}{n}}}
-\frac{2\beta^{2}}{(nDt+c_{2})^{\frac{2}{n}}}-\frac{c_{0}}{(nDt+c_{2})^{\frac{3}{n}}}}
{\frac{3D^{2}}{(nDt+c_{2})^{2}}-\frac{l^{2}}{(nDt+c_{2})^{\frac{6}{n}}}
-\frac{3\beta^{2}}{(nDt+c_{2})^{\frac{2}{n}}}-\frac{c_{0}}{(nDt+c_{2})^{\frac{3}{n}}}}.
\end{equation}
Then, using the equation (4) we find the Tachyon potential as follow,
\begin{eqnarray}\label{s24}
V^{2}&=&\frac{2nD^{2}l^{2}}{(nDt+c_{2})^{\frac{6+2n}{n}}}+\frac{6(n-2)\beta^{2}D^{2}}{(nDt+c_{2})^{\frac{2+2n}{n}}}
+\frac{c_{0}(2n-3)D^{2}}{(nDt+c_{2})^{\frac{3+2n}{n}}}-\frac{3(2n-3)D^{4}}{(nDt+c_{2})^{4}}\nonumber\\
&-&\frac{l^{2}}{(nDt+c_{2})^{\frac{12}{n}}}-\frac{c_{0}l^{2}}{(nDt+c_{2})^{\frac{9}{n}}}
-\frac{2\beta^{2}l^{2}}{(nDt+c_{2})^{\frac{8}{n}}}+\frac{c_{0}^{2}\beta^{2}}{(nDt+c_{2})^{\frac{5}{n}}}
+\frac{3\beta^{4}}{(nDt+c_{2})^{\frac{4}{n}}}.
\end{eqnarray}
It is clear that the Tachyon potential vanished at the late time.\\
In Fig. 2 we find behavior of the scalar field in terms of time which is increasing function. We vary $n$ and find that increasing $n$ increases value of the scalar field.\\
In Fig. 3 we give plot of total EoS for various values of $n$. For the selected values $0<n\leq1$ we obtain $-1<\omega\leq-1/3$ after initial time. We find that $n>1$ yield to positive EoS at the late time. At the early universe we can see that $\omega<-1$. It means that infinitesimal $n$ yields to phantom like universe while large $n$ yields to quintessence like universe.

\begin{figure}[th]
\begin{center}
\includegraphics[scale=.35]{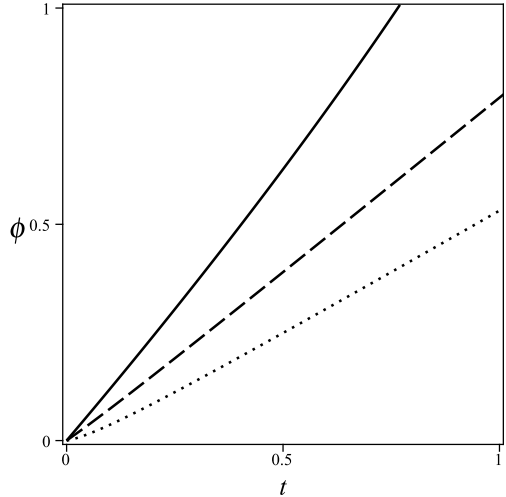}
\caption{Plot of scalar field in terms of cosmic time with $c_{0}=2$, $l=1.2$, $c_{2}=1$, $D=2.5$ and $\beta=0.7$. $n=2$ (solid line), $n=1$ (dashed line), $n=0.5$ (dotted line).}
\end{center}
\end{figure}

\begin{figure}[th]
\begin{center}
\includegraphics[scale=.35]{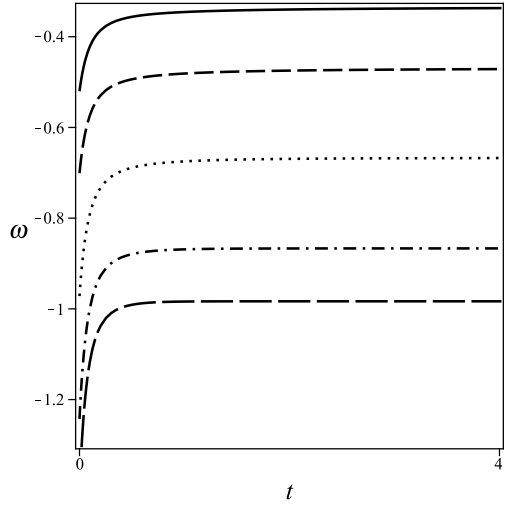}
\caption{Plot of EoS parameter in terms of cosmic time with $c_{0}=2$, $l=1.2$, $c_{2}=1$, $D=2.5$ and $\beta=0.7$. $n=1$ (solid line), $n=0.8$ (dashed line), $n=0.5$ (dotted line), $n=0.2$ (dash dotted line), $n=0.025$ (long dashed line).}
\end{center}
\end{figure}

\subsection{Barotropic matter}
We can consider barotropic matter with constant $\omega^{(m)}$ as matter contribution.

\subsubsection{$n=0$}

In that case one can obtain,
\begin{equation}\label{s25}
\rho^{(m)}=c_{0}c_{1}^{-3(1+\omega^{(m)})}e^{-3(1+\omega^{(m)})Dt},
\end{equation}
which gives us the following relation,
\begin{equation}\label{s26}
\dot{\phi}^{2}=\frac{2\frac{l^{2}}{c_{1}^{6}}e^{-6Dt}+2\frac{\beta^{2}}{c_{1}^{2}}e^{-2Dt}+c_{0}c_{1}^{-3(1+\omega^{(m)})}e^{-3(1+\omega^{(m)})Dt}}
{\frac{l^{2}}{c_{1}^{6}}e^{-6Dt}+3\frac{\beta^{2}}{c_{1}^{2}}e^{-2Dt}+c_{0}c_{1}^{-3(1+\omega^{(m)})}e^{-3(1+\omega^{(m)})Dt}-3D^{2}}.
\end{equation}
Then, using the equation (4) we find the Tachyon potential as follow,
\begin{eqnarray}\label{s27}
V^{2}=9D^{4}&-&12D^{2}\frac{\beta^{2}}{c_{1}^{2}}e^{-2Dt}-3D^{2}c_{0}c_{1}^{-3(1+\omega^{(m)})}e^{-3(1+\omega^{(m)})Dt}\nonumber\\
&+&3\frac{\beta^{4}}{c_{1}^{4}}e^{-4Dt}+c_{0}\frac{\beta^{2}}{c_{1}^{2}}c_{1}^{-3(1+\omega^{(m)})}e^{-(5+3\omega^{(m)})Dt}\nonumber\\
&-&2\frac{l^{2}\beta^{2}}{c_{1}^{8}}e^{-8Dt}-c_{0}\frac{l^{2}}{c_{1}^{6}}c_{1}^{-3(1+\omega^{(m)})}e^{-3(3+\omega^{(m)})Dt}-\frac{l^{4}}{c_{1}^{12}}e^{-12Dt}.
\end{eqnarray}
It is clear that the Tachyon potential yields to $3D^{2}$ at the late time which is similar to the pressureless matter. Also, Fig. 4 show similar behavior of the scale factor with the previous model. It increases with time and also increases with $D$.

\begin{figure}[th]
\begin{center}
\includegraphics[scale=.35]{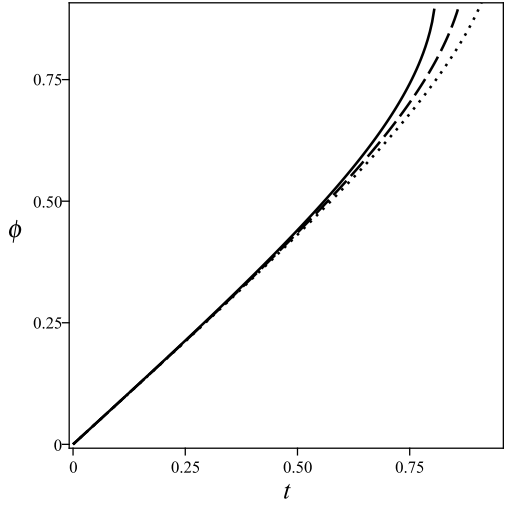}
\caption{Plot of scalar field in terms of cosmic time with $c_{0}=2$, $c_{1}=1$, $\omega^{(m)}=-0.5$, $l=\sqrt{5}$ and $\beta=10$. $D=2$ (solid line), $D=1.9$ (dashed line), $D=1.8$ (dotted line).}
\end{center}
\end{figure}

\subsubsection{$n\neq0$}

In that case one can obtain,
\begin{equation}\label{s28}
\rho^{(m)}=c_{0}(nDt+c_{2})^{-\frac{3(1+\omega^{(m)})}{n}}.
\end{equation}
Similar to the previous, we can find,
\begin{equation}\label{s29}
\dot{\phi}^{2}=\frac{\frac{2nD^{2}}{(nDt+c_{2})^{2}}-\frac{2l^{2}}{(nDt+c_{2})^{\frac{6}{n}}}
-\frac{2\beta^{2}}{(nDt+c_{2})^{\frac{2}{n}}}-\frac{c_{0}(1+\omega^{(m)})}{(nDt+c_{2})^{\frac{3(1+\omega^{(m)})}{n}}}}
{\frac{3D^{2}}{(nDt+c_{2})^{2}}-\frac{l^{2}}{(nDt+c_{2})^{\frac{6}{n}}}
-\frac{3\beta^{2}}{(nDt+c_{2})^{\frac{2}{n}}}-\frac{c_{0}}{(nDt+c_{2})^{\frac{3(1+\omega^{(m)})}{n}}}}.
\end{equation}
Then, using the equation (4) we find the Tachyon potential as follow,
\begin{eqnarray}\label{s30}
V^{2}&=&\frac{2nD^{2}l^{2}}{(nDt+c_{2})^{\frac{6+2n}{n}}}+\frac{6(n-2)\beta^{2}D^{2}}{(nDt+c_{2})^{\frac{2+2n}{n}}}
+\frac{c_{0}(2n-3)D^{2}}{(nDt+c_{2})^{\frac{3(1+\omega^{(m)})+2n}{n}}}-\frac{3(2n-3)D^{4}}{(nDt+c_{2})^{4}}\nonumber\\
&-&\frac{l^{2}}{(nDt+c_{2})^{\frac{12}{n}}}-\frac{c_{0}l^{2}}{(nDt+c_{2})^{\frac{3(3+\omega^{(m)})}{n}}}
-\frac{2\beta^{2}l^{2}}{(nDt+c_{2})^{\frac{8}{n}}}+\frac{c_{0}\beta^{2}}{(nDt+c_{2})^{\frac{5+3\omega^{(m)}}{n}}}\nonumber\\
&+&\frac{3\beta^{4}}{(nDt+c_{2})^{\frac{4}{n}}}
+\frac{3c_{0}D^{2}\omega^{(m)}}{(nDt+c_{2})^{\frac{3(1+\omega^{(m)})+2n}{n}}}
-\frac{l^{2}c_{0}\omega^{(m)}}{(nDt+c_{2})^{\frac{3(3+\omega^{(m)})}{n}}}\nonumber\\
&-&\frac{3\beta^{2}c_{0}\omega^{(m)}}{(nDt+c_{2})^{\frac{5+3\omega^{(m)}}{n}}}
-\frac{c_{0}^{2}\omega^{(m)}}{(nDt+c_{2})^{\frac{2(4+3\omega^{(m)})}{n}}}.
\end{eqnarray}
It is clear that the potential vanishes at the late time which is expected for tachyon potential.\\
Numerically, we find behavior of the scaler field in Fig. 5. It is illustrated that increasing $n$ increases value of the scaler field. Comparing with the previous model with pressureless matter we obtain enhanced scalar field.\\
Also, Fig. 6 shows variation of EoS parameter similar to the previous model. For the selected values $0<n<1$ we obtain $-1<\omega\leq-1/3$ after initial time (initially we have $\omega<-1$). It is obvious that $n>1$ yields to positive EoS after some time. It means that infinitesimal $n$ yields to phantom like universe while large $n$ yields to quintessence like universe. For some cases with $0<n<1$ we can see transition form phantom to quintessence like universes.\\

\begin{figure}[th]
\begin{center}
\includegraphics[scale=.35]{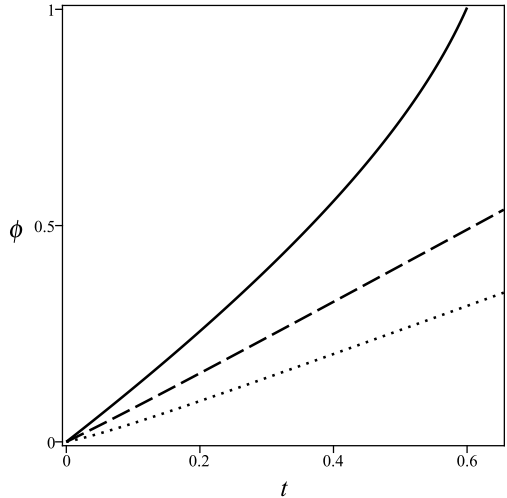}
\caption{Plot of scalar field in terms of cosmic time with $c_{0}=2$, $l=1.2$, $c_{2}=1$, $D=2.5$, $\omega^{(m)}=-0.5$ and $\beta=0.7$. $n=2$ (solid line), $n=1$ (dashed line), $n=0.5$ (dotted line).}
\end{center}
\end{figure}

\begin{figure}[th]
\begin{center}
\includegraphics[scale=.35]{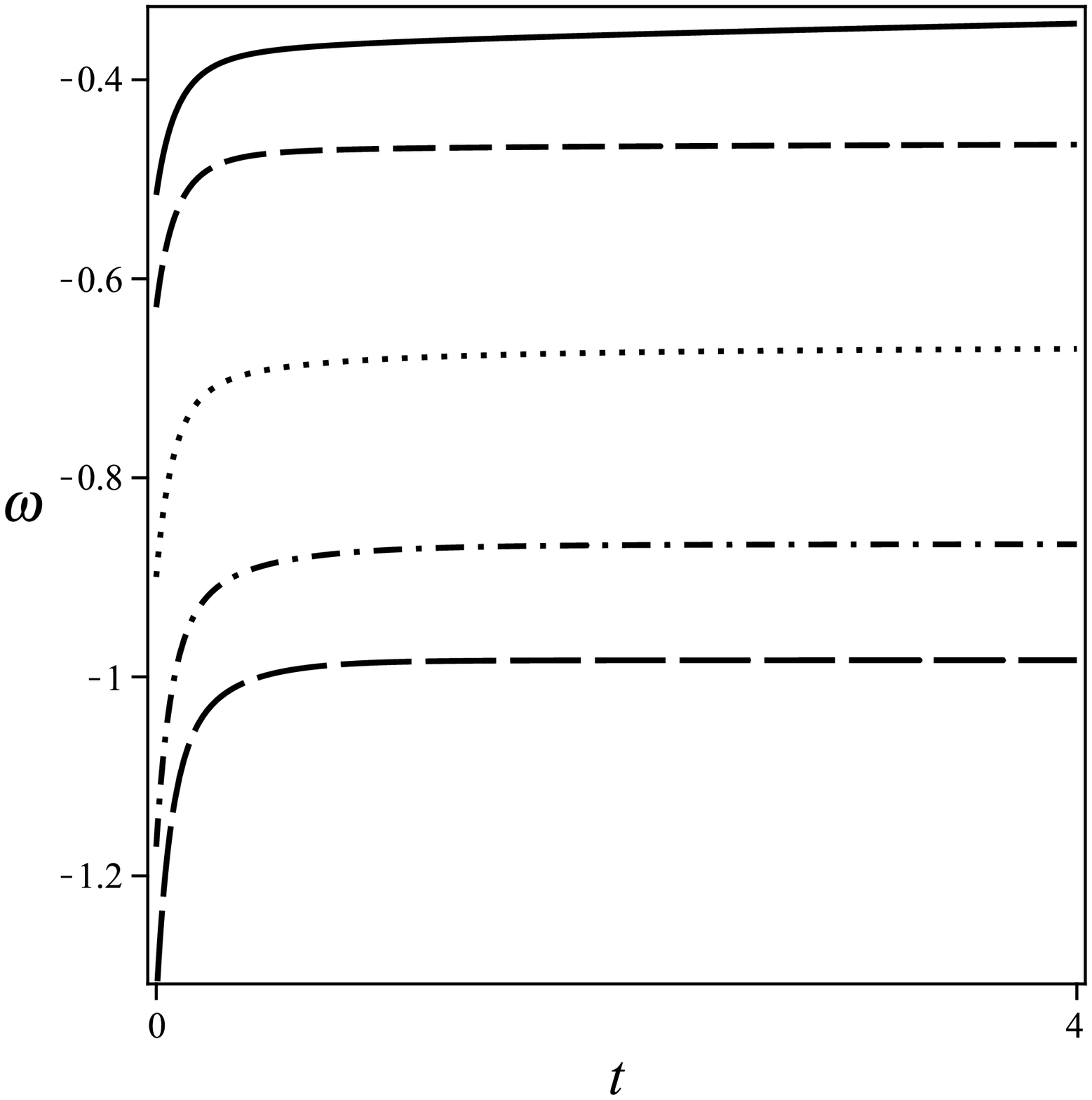}
\caption{Plot of EoS parameter in terms of cosmic time with $c_{0}=2$, $l=1.2$, $c_{2}=1$, $D=2.5$, $\omega^{(m)}=-0.5$ and $\beta=0.7$. $n=0.925$ (solid line), $n=0.8$ (dashed line), $n=0.5$ (dotted line), $n=0.2$ (dash dotted line), $n=0.025$ (long dashed line).}
\end{center}
\end{figure}

\subsection{Modified Chaplygin gas}
In the last model we can consider modified Chaplygin gas as a matter with the following energy density.
\begin{equation}\label{s31}
\rho_{mcg}=\left(\frac{\mathcal{B}}{1+\gamma}+\frac{\mathcal{C}}{a^{3(1+\gamma)(1+\alpha)}}\right)^{\frac{1}{1+\alpha}},
\end{equation}
with,
\begin{equation}\label{s32}
\mathcal{C}=\frac{\mathcal{B}}{1+\gamma}a(0)^{3(1+\gamma)(1+\alpha)},
\end{equation}
as an integration constant where $a$ is scale factor with current constant value $a(0)$.\\
We can use the above matter density in the equation (14) to obtain dark energy density.
\subsubsection{$n=0$}
\begin{equation}\label{s33}
\dot{\phi}^{2}=\frac{2\frac{l^{2}}{c_{1}^{6}}e^{-6Dt}+2\frac{\beta^{2}}{c_{1}^{2}}e^{-2Dt}
+\left(\frac{\mathcal{B}}{1+\gamma}+\frac{\mathcal{C}}{(c_{1}e^{Dt})^{3(1+\gamma)(1+\alpha)}}\right)^{\frac{1}{1+\alpha}}}
{\frac{l^{2}}{c_{1}^{6}}e^{-6Dt}+3\frac{\beta^{2}}{c_{1}^{2}}e^{-2Dt}
+\left(\frac{\mathcal{B}}{1+\gamma}+\frac{\mathcal{C}}{(c_{1}e^{Dt})^{3(1+\gamma)(1+\alpha)}}\right)^{\frac{1}{1+\alpha}}-3D^{2}}.
\end{equation}
Then, we find the Tachyon potential as follow,
\begin{eqnarray}\label{s34}
V^{2}&=&9D^{4}-12D^{2}\frac{\beta^{2}}{c_{1}^{2}}e^{-2Dt}
-3D^{2}\left(\frac{\mathcal{B}}{1+\gamma}+\frac{\mathcal{C}}{(c_{1}e^{Dt})^{3(1+\gamma)(1+\alpha)}}\right)^{\frac{1}{1+\alpha}}\nonumber\\
&+&3\frac{\beta^{4}}{c_{1}^{4}}e^{-4Dt}
+\left(\frac{\mathcal{B}}{1+\gamma}+\frac{\mathcal{C}}{(c_{1}e^{Dt})^{3(1+\gamma)(1+\alpha)}}\right)^{\frac{1}{1+\alpha}}\frac{\beta^{2}}{c_{1}^{2}}e^{-2Dt}
-\frac{l^{4}}{c_{1}^{12}}e^{-12Dt}\nonumber\\
&-&2\frac{l^{2}\beta^{2}}{c_{1}^{8}}e^{-8Dt}
-\left(\frac{\mathcal{B}}{1+\gamma}+\frac{\mathcal{C}}{(c_{1}e^{Dt})^{3(1+\gamma)(1+\alpha)}}\right)^{\frac{1}{1+\alpha}}\frac{l^{2}}{c_{1}^{6}}e^{-6Dt}.
\end{eqnarray}
Similar to the previous models the Tachyon potential yields to a constant at the late time.\\
In Fig. 7 we find behavior of the scalar field numerically which has not main differences with previous models.
\begin{figure}[th]
\begin{center}
\includegraphics[scale=.35]{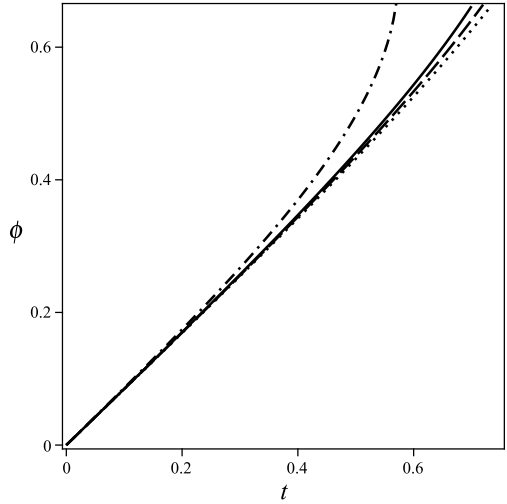}
\caption{Plot of scalar field in terms of cosmic time with $\mathcal{C}=1$, $c_{1}=1$, $l=\sqrt{5}$, $\gamma=0.3$, $\alpha=0.5$ and $\beta=10$.$D=2.5$ (dash dotted line), $D=2$ (solid line), $D=1.9$ (dashed line), $D=1.8$ (dotted line).}
\end{center}
\end{figure}

\subsubsection{$n\neq0$}
In that case we can obtain,
\begin{equation}\label{s35}
\dot{\phi}^{2}=\frac{\frac{2nD^{2}}{(nDt+c_{2})^{2}}-\frac{2l^{2}}{(nDt+c_{2})^{\frac{6}{n}}}
-\frac{2\beta^{2}}{(nDt+c_{2})^{\frac{2}{n}}}
-\omega\left(\frac{\mathcal{B}}{1+\gamma}+\frac{\mathcal{C}}{(nDt+c_{2})^{\frac{3(1+\gamma)(1+\alpha)}{n}}}\right)^{\frac{1}{1+\alpha}}}
{\frac{3D^{2}}{(nDt+c_{2})^{2}}-\frac{l^{2}}{(nDt+c_{2})^{\frac{6}{n}}}
-\frac{3\beta^{2}}{(nDt+c_{2})^{\frac{2}{n}}}
-\left(\frac{\mathcal{B}}{1+\gamma}+\frac{\mathcal{C}}{(nDt+c_{2})^{\frac{3(1+\gamma)(1+\alpha)}{n}}}\right)^{\frac{1}{1+\alpha}}},
\end{equation}
where $\omega$ given by the equation (5).
Then, we can find the Tachyon potential as follow,
\begin{eqnarray}\label{s36}
V^{2}&=&\frac{2nD^{2}l^{2}}{(nDt+c_{2})^{\frac{6+2n}{n}}}+\frac{6(n-2)\beta^{2}D^{2}}{(nDt+c_{2})^{\frac{2+2n}{n}}}
-\frac{3(2n-3)D^{4}}{(nDt+c_{2})^{4}}-\frac{l^{2}}{(nDt+c_{2})^{\frac{12}{n}}}\nonumber\\
&+&\frac{(2n-3(1-\omega))D^{2}
\left(\frac{\mathcal{B}}{1+\gamma}+\frac{\mathcal{C}}{(nDt+c_{2})^{\frac{3(1+\gamma)(1+\alpha)}{n}}}\right)
^{\frac{1}{1+\alpha}}}{(nDt+c_{2})^{2}}-\frac{2\beta^{2}l^{2}}{(nDt+c_{2})^{\frac{8}{n}}}\nonumber\\
&-&\frac{\left(\frac{\mathcal{B}}{1+\gamma}+\frac{\mathcal{C}}{(nDt+c_{2})^{\frac{3(1+\gamma)(1+\alpha)}{n}}}\right)^{\frac{1}{1+\alpha}}(1+\omega)l^{2}}{(nDt+c_{2})^{\frac{6}{n}}}
-\frac{(3\omega-1)\beta^{2}\left(\frac{\mathcal{B}}{1+\gamma}
+\frac{\mathcal{C}}{(nDt+c_{2})^{\frac{3(1+\gamma)(1+\alpha)}{n}}}\right)^{\frac{1}{1+\alpha}}}{(nDt+c_{2})^{\frac{2}{n}}}\nonumber\\
&-&\omega\left(\frac{\mathcal{B}}{1+\gamma}
+\frac{\mathcal{C}}{(nDt+c_{2})^{\frac{3(1+\gamma)(1+\alpha)}{n}}}\right)^{\frac{2}{1+\alpha}}+\frac{3\beta^{4}}{(nDt+c_{2})^{\frac{4}{n}}},
\end{eqnarray}
which vanished at the late time as expected.\\
Time evolution of the scalar field illustrated in Fig. 8 in agreement with previous models. However we can see new manner for the EoS parameter in Fig. 9. For some values of $0<n\leq1$ the EoS parameter is in the range of $\omega\geq-1$ initially. After late time we have $\omega<-1$ which is characteristic of phantom universe. It means that the universe transfer from quintessence at the early time to phantom at the late time.

\begin{figure}[th]
\begin{center}
\includegraphics[scale=.35]{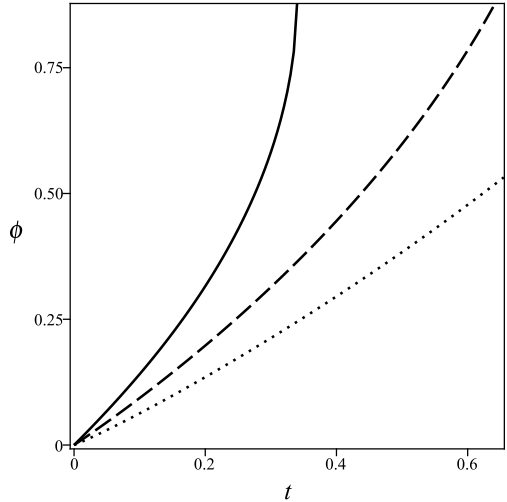}
\caption{Plot of scalar field in terms of cosmic time with $c_{2}=1$, $l=1.2$, $\mathcal{C}=1$, $\gamma=0.3$, $\alpha=0.3$, $D=2.5$, and $\beta=0.7$. $n=2$ (solid line), $n=1$ (dashed line), $n=0.5$ (dotted line).}
\end{center}
\end{figure}

\begin{figure}[th]
\begin{center}
\includegraphics[scale=.35]{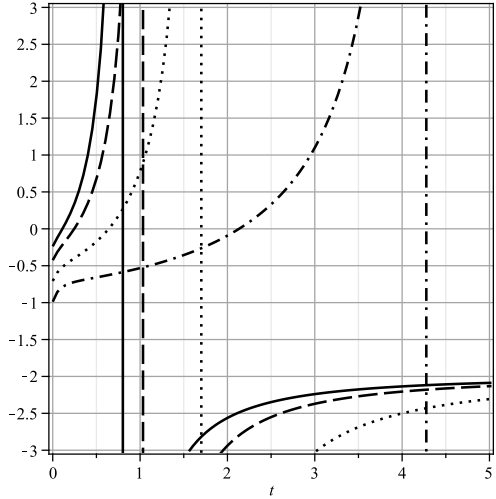}
\caption{Plot of EoS parameter in terms of cosmic time with $c_{2}=1$, $l=1.2$, $\mathcal{C}=1$, $\gamma=0.3$, $\alpha=0.3$, $D=2.5$, and $\beta=0.7$. $n=1$ (solid line), $n=0.8$ (dashed line), $n=0.5$ (dotted line), $n=0.2$ (dash dotted line).}
\end{center}
\end{figure}

\section{Stability}
We can investigate stability of our models by using analysis of sound speed,
\begin{equation}\label{s37}
V_{s}^{2}=\frac{\dot{p}^{(de)}}{\dot{\rho}^{(de)}}.
\end{equation}
We performed numerical analysis for three models. Fig. 10 shows our results for the case of pressureless matter. We can see that the model may be stable after initial time for $n=1.5$ and $D\geq2$. There are also some situation where the model is stable initially but instable after the late time.\\
plots of Fig . 11 show our numerical results for barotropic matter with constant EoS. We find that stability of this model is depend on $D$, $n$ and $\omega^{(m)}$. Having $D\geq2$, $n\geq1.5$, and $0>\omega^{(m)}>-1$ give completely stable model. However the special case of $n=0$ yields to positive constant sound speed which guarantee stability of the model.\\
Finally in the case of modified Chaplygin gas we find that the model is initially instable which will be stable after some time for $D\geq2$, $n\geq1.5$ (see Fig. 12). In this case also we have constant positive sound speed corresponding to the $n=0$.\\
Comparing the $n=0$ case of squared sound speed for the modified Chaplygin gas and barotropic model tells that the $V_{s}^{2}$ has smaller value in modified Chaplygin gas model.\\
In all model we can have completely stable model by choosing $D\geq2$, $n>1.5$ and EoS parameter corresponding to quintessence like universe.

\begin{figure}[th]
\begin{center}
\includegraphics[scale=.35]{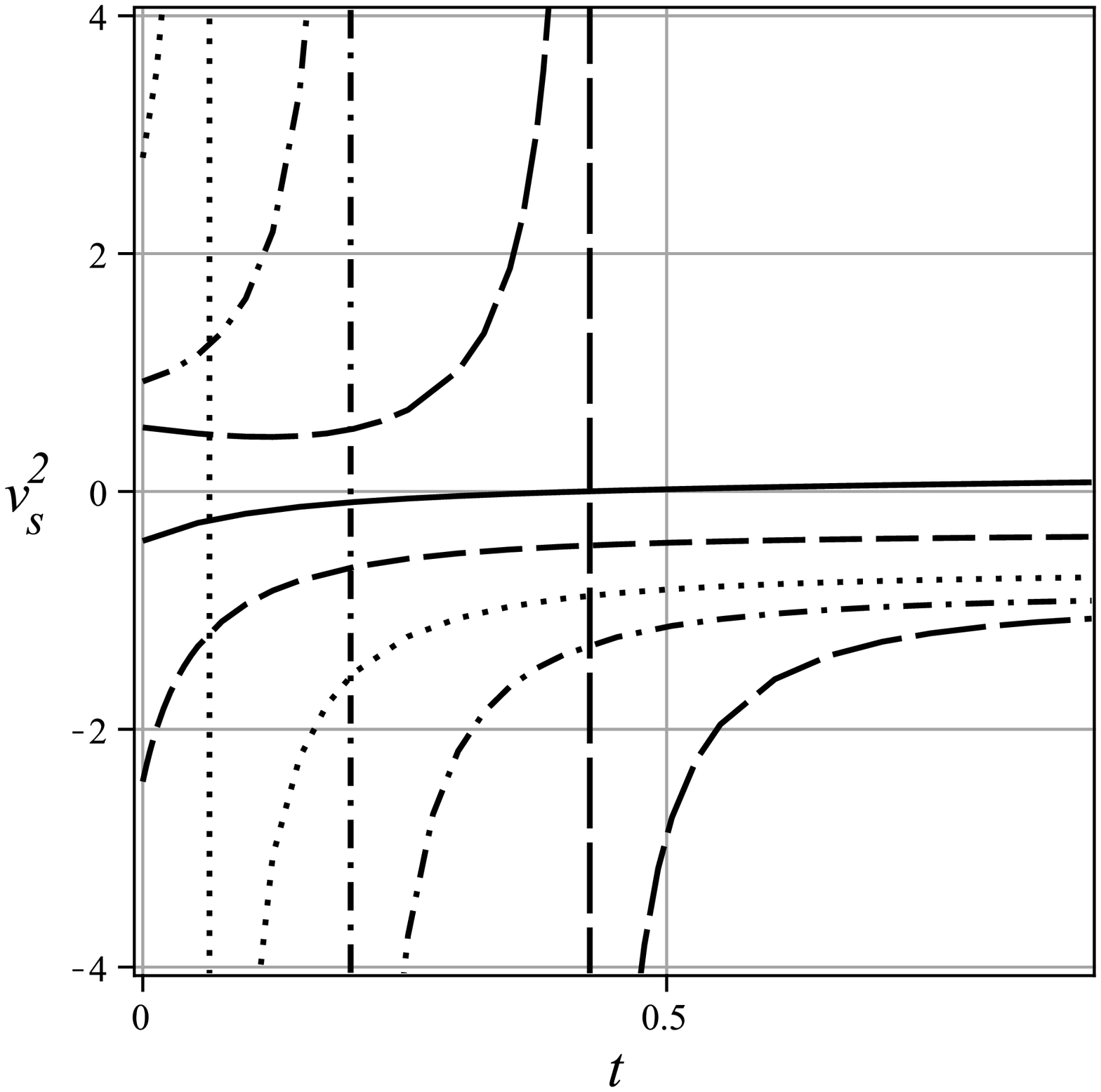}\includegraphics[scale=.35]{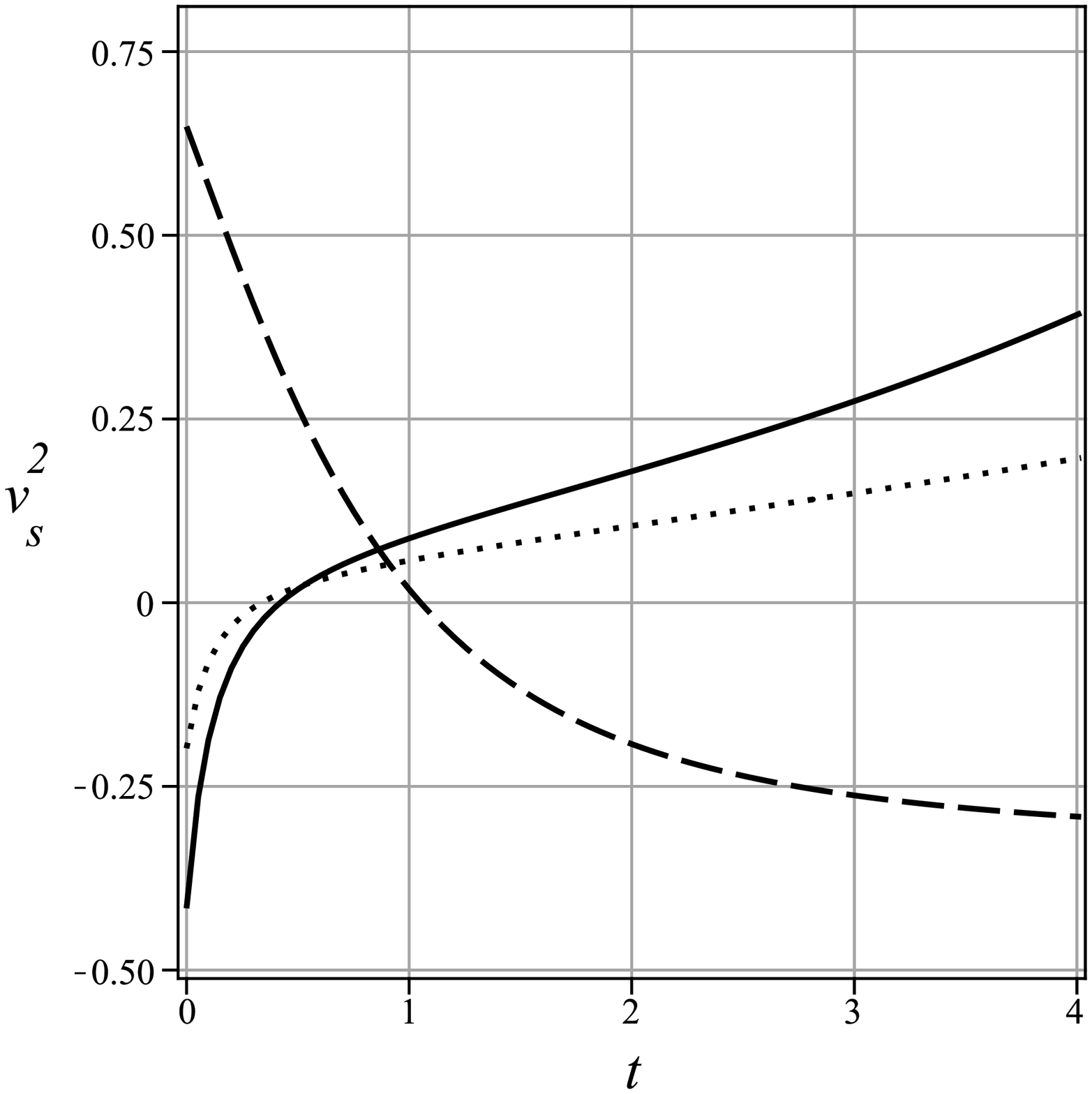}
\caption{Squared sound speed in terms of cosmic time with $c_{1}=1$, $l=1.2$, $c_{0}=1$, and $\beta=0.7$. Left: $D=2$, $n=1.5$ (solid line), $n=1$ (dashed line), $n=0.5$ (dotted line), $n=0.2$ (dash dotted line), $n=0.05$ (long dashed line). Right: $n=1.5$, $D=2$ (solid line), $D=0.8$ (dashed line), $D=2.5$ (dotted line).}
\end{center}
\end{figure}

\begin{figure}[th]
\begin{center}
\includegraphics[scale=.27]{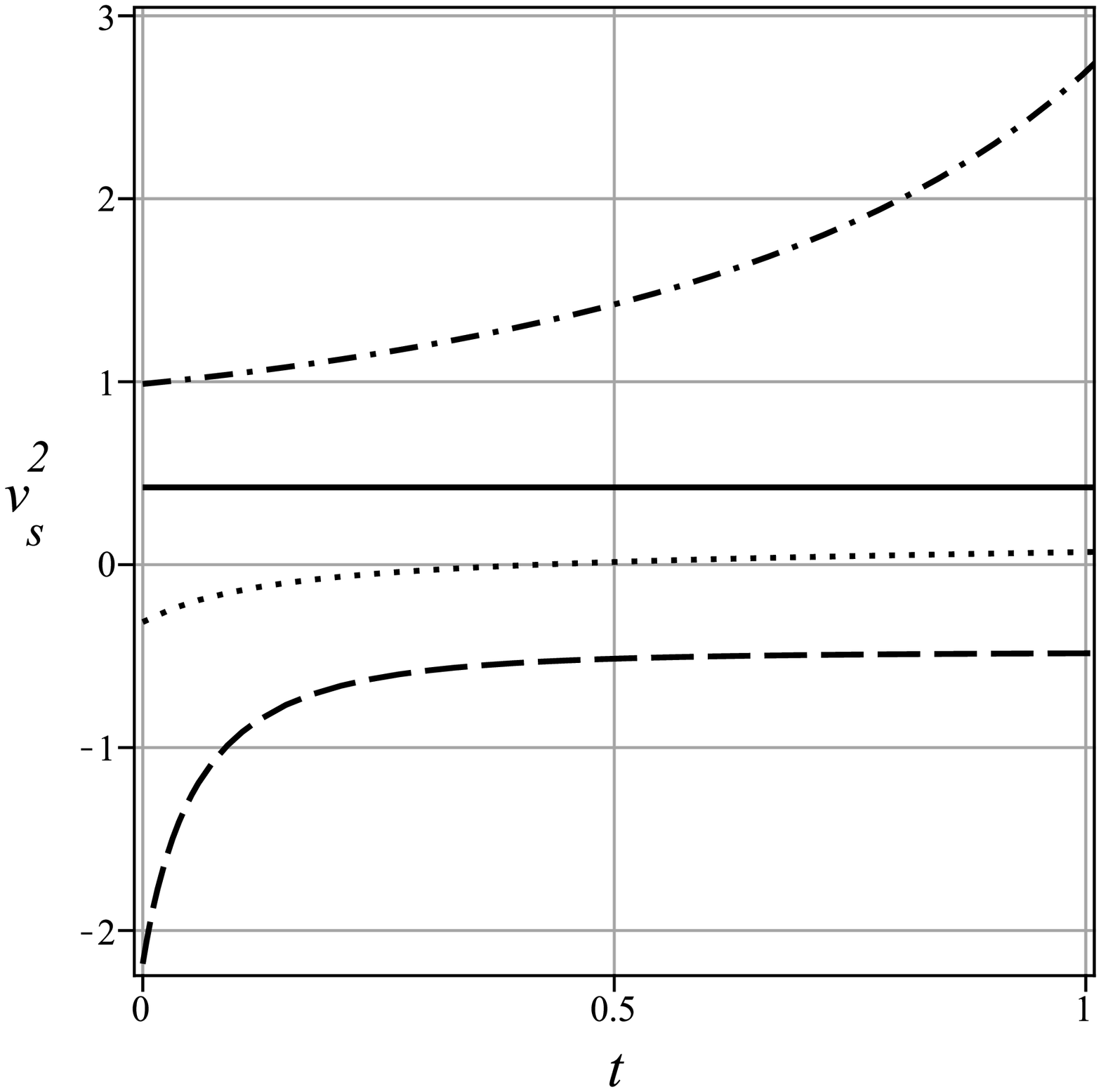}\includegraphics[scale=.27]{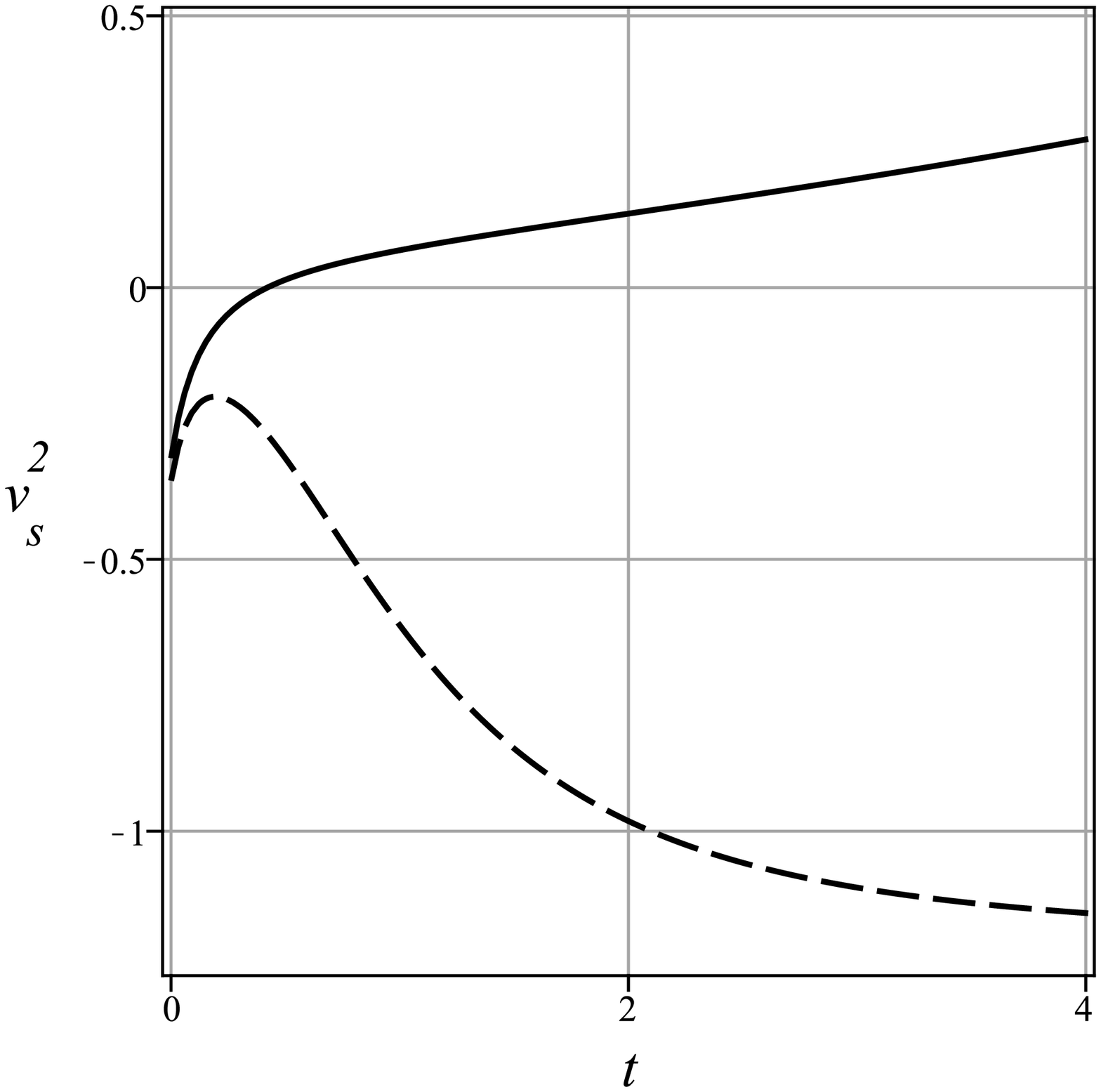}\includegraphics[scale=.27]{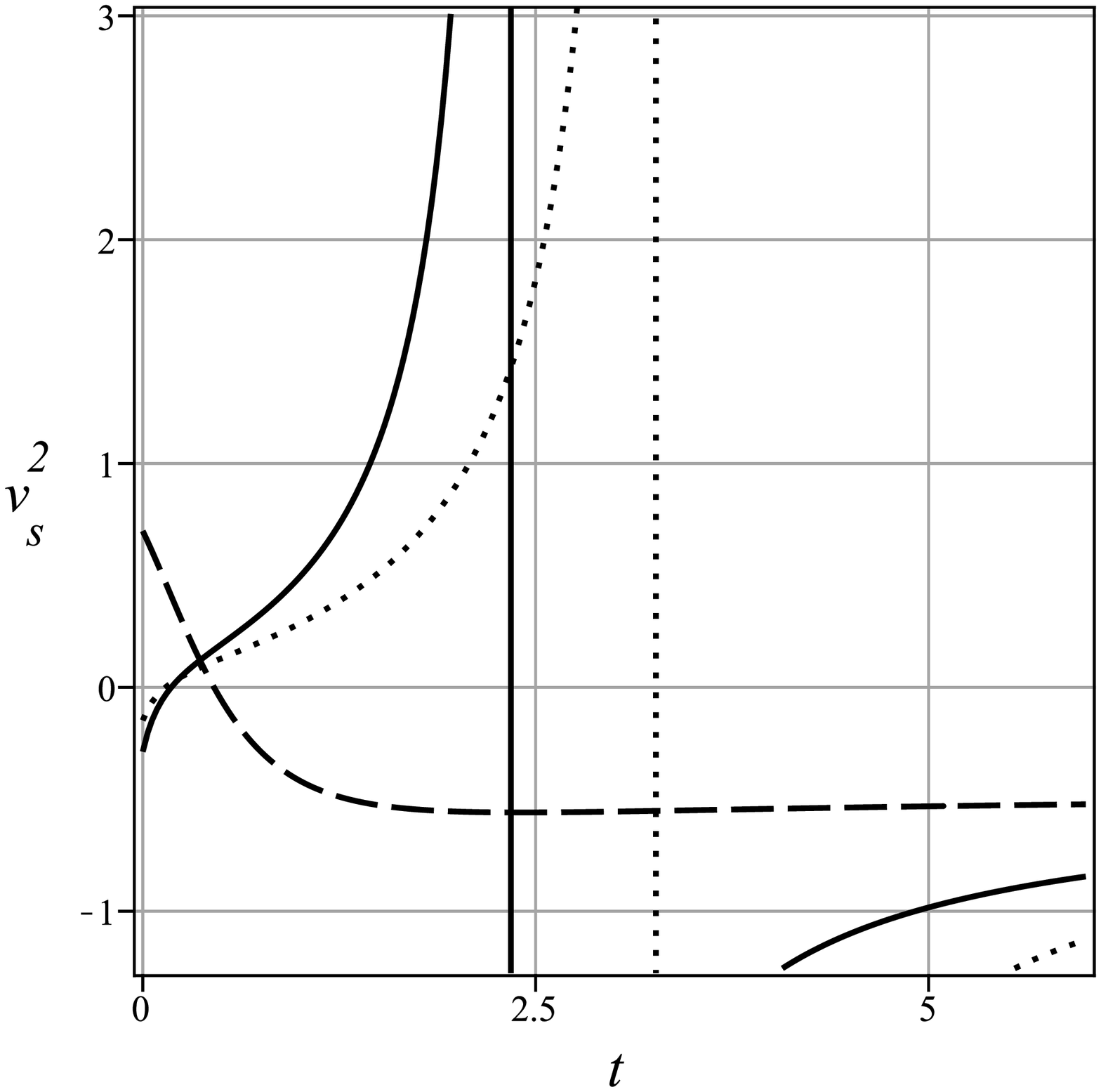}
\caption{Squared sound speed in terms of cosmic time with $c_{2}=1$, $l=1.2$, $c_{0}=1$, and $\beta=0.7$. Left: $D=2$, $\omega^{(m)}=-0.5$, $n=0$ (solid line), $n=0.8$ (dashed line), $n=1.5$ (dotted line), $n=2.9$ (dash dotted line). Middle: $D=2$, $n=1.5$, $0>\omega^{(m)}>-1$ (solid line), $\omega^{(m)}<-1$ (dashed line). Right: $n=1.5$, $\omega^{(m)}=-0.5$, $D=2$ (solid line), $D=0.8$ (dashed line), $D=2.5$ (dotted line).}
\end{center}
\end{figure}

\begin{figure}[th]
\begin{center}
\includegraphics[scale=.35]{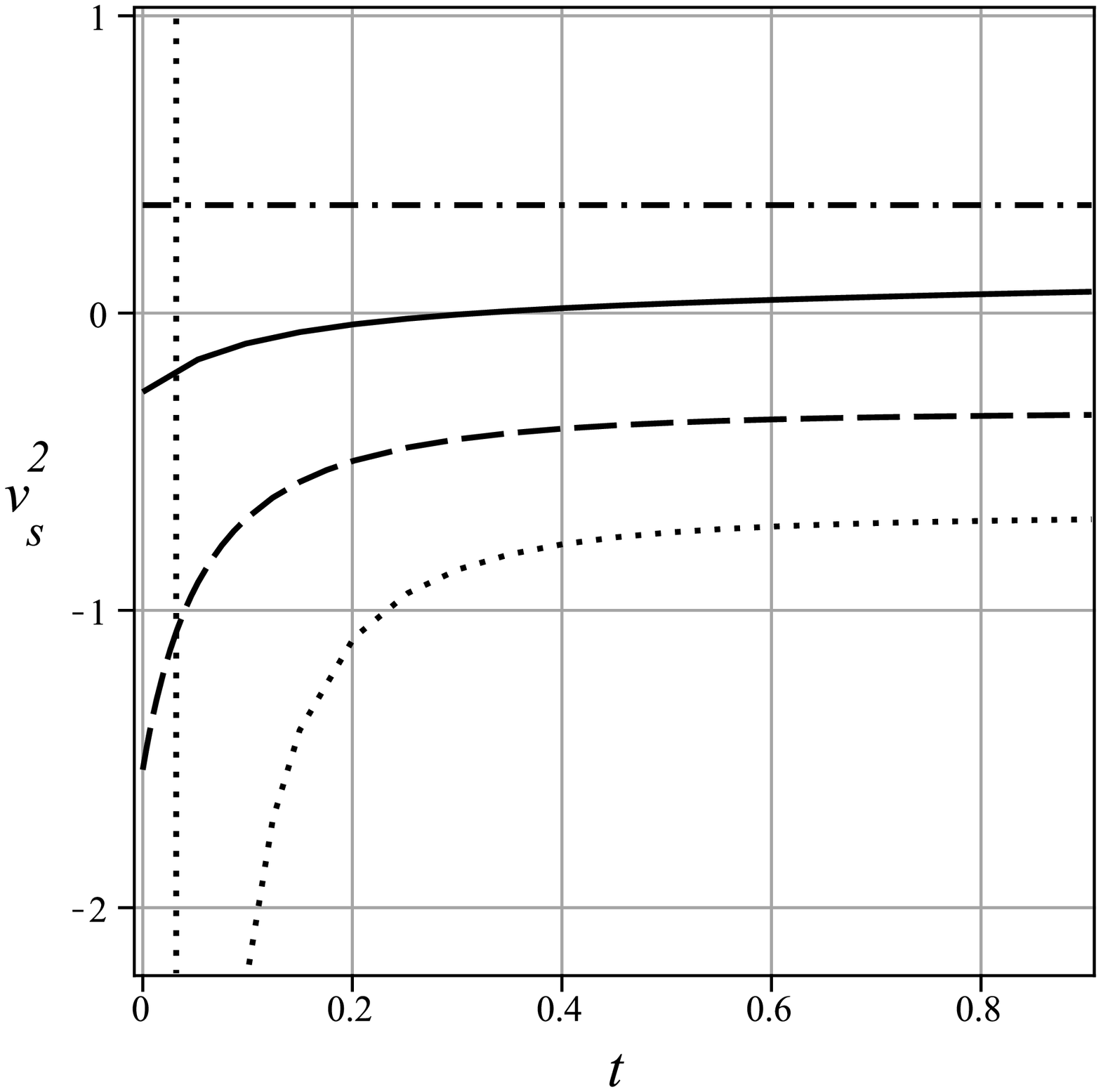}\includegraphics[scale=.35]{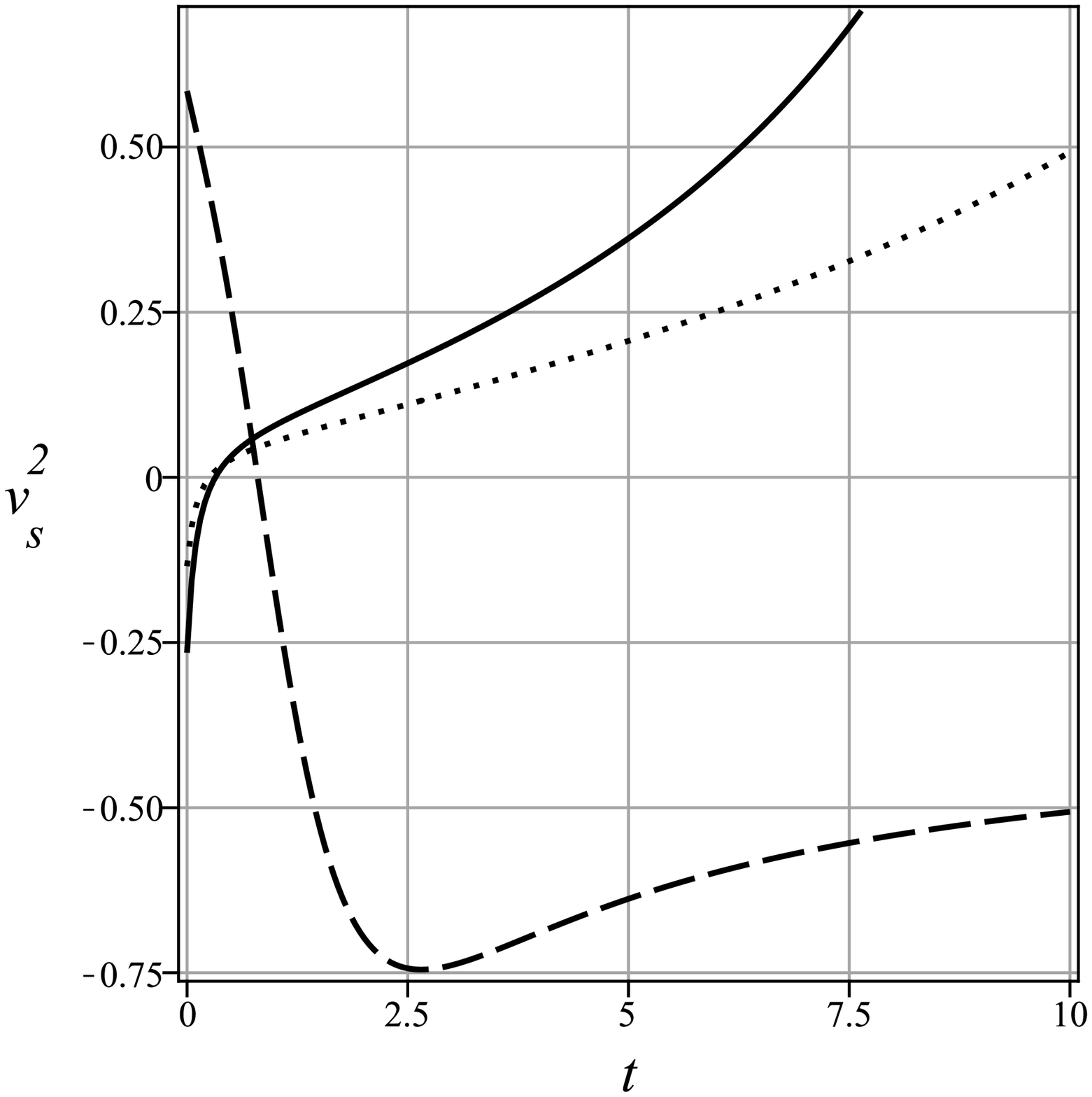}
\caption{Squared sound speed in terms of cosmic time with $c_{2}=1$, $l=1.2$, $\mathcal{C}=1$, $\gamma=0.3$, $\alpha=0.3$, and $\beta=0.7$. Left: $D=2$, $n=1.5$ (solid line), $n=1$ (dashed line), $n=0.5$ (dotted line), $n=0$ (dash dotted line). Right: $n=1.5$, $D=2$ (solid line), $D=0.8$ (dashed line), $D=2.5$ (dotted line).}
\end{center}
\end{figure}

\section{Conclusion}
In this work, we proposed a cosmological model to describe universe in Bianchi Type-V based on Tachyon scalar field. Indeed, we considered Tachyon scalar field as dark energy and three different types as matter. We neglected, at this step, interaction between them which may be considered in future study. In the first model we considered pressureless matter, in the second model we considered barotropic matter with constant EoS parameter and in the third model we considered modified Chaplygin gas. Under assumption of $H=Da^{-1}$ we obtained analytical expressions for the scale factor, dark energy densities, $\dot{\phi} $ and scalar potential $V$ in different cases of $n=0$ and $n\neq0$. Therefore, we could discuss numerically about evolution of the scalar field in all models. We found that the Tachyon potential vanishes at the late time for the case of $n\neq0$ as expected while yields to a constant proportional to $D$ for the case of $n=0$.\\
We didn't find main difference between the scalar fields of pressureless matter and barotropic matter. But we found that the value of the scalar field in the third model is smaller than other models (comparing Figs. 1, 4 and 7). Also we found that the variation of $D$ is not important at the initial time, but it increases value of the scalar field at the late time.\\
In the case of $n\neq0$ we found that the value of the scalar field in the third model is bigger than the second model and in the second model is bigger than the first model (comparing Figs. 2, 5 and 8). It means that the modified Chaplygin gas model has larger scalar field and pressureless matter has smallest scalar field during time. We found that, value of $n$ increases value of the scalar field in all models.\\
In the case of $n\neq0$ we also discussed about total equation of state parameter numerically. In the first model we found that, for the selected values of $n$ in the range of $0 < n \leq 1$
we obtained $-1 < \omega\leq -1/3$ after initial time. We found that $n > 1$ yield to positive EoS at
the late time. Also at the early universe with $n\ll1$ we have shown that $\omega<-1$, which means that infinitesimal values of $n$ yields to phantom like universe while large $n$ yields to quintessence like universe.\\
In the second model we found that, for the selected
values of $n$ in the range of $0 < n < 1$ we have $-1 < \omega\leq -1/3$ just before.
It has been shown that $n > 1$ yields to positive EoS after some time. It means that infinitesimal
$n$ yields to phantom like universe while large $n$ yields to quintessence like universe. For
some cases with $0 < n < 1$ we can see transition form phantom to quintessence like universes.\\
Finally in the third model, we found that for some values of $n$ in the range of $0 < n \leq 1$ the EoS parameter is in the range of $\omega\geq-1$ at the early universe. After late time we have
$\omega<-1$ which is characteristic of phantom universe. It means that the universe transfer
from quintessence at the early time to phantom at the late time. Comparing to the first and second model we can see opposite behavior. In the first and second model we obtain current universe as quintessence like, while in the case of modified Chaplygin gas the universe is in phantom phase today.\\
Finally we found that all models may be completely stable for $D\geq2$, $n>1.5$. However, we can have stable models at present for other values. Just there are some instability at the early universe which may cause of some particle creation at that step.\\
Here, there are several interesting problem which left for future works. For example we can investigate possibility of varying $G$ and $\Lambda$ [37] or interaction between components [38, 39]. In these case one may obtain more agreement with observational data.

\end{document}